\documentclass[conference]{IEEEtran}
\IEEEoverridecommandlockouts

\usepackage{cite}
\usepackage{amsmath,amssymb,amsfonts}
\usepackage{algorithmic}
\usepackage[ruled,linesnumbered]{algorithm2e}
\usepackage{graphicx}
\usepackage{textcomp}
\usepackage[colorlinks]{hyperref}
\hypersetup{citecolor=blue}
\usepackage{xcolor}
\usepackage{threeparttable}
\def\BibTeX{{\rm B\kern-.05em{\sc i\kern-.025em b}\kern-.08em
    T\kern-.1667em\lower.7ex\hbox{E}\kern-.125emX}}
\begin{document}
\title{Low-Complexity CSI Acquisition Exploiting Geographical Diversity in Fluid Antenna System}
\author{
	\IEEEauthorblockN{
		Zhentian Zhang\textsuperscript{1}, David Morales-Jimenez\textsuperscript{3}, 
		Jian Dang\textsuperscript{1,2}, \\
		Zaichen Zhang\textsuperscript{1,2},
		Christos Masouros\textsuperscript{5},
		Hao Jiang\textsuperscript{4}
		}
	\IEEEauthorblockA{
		\textsuperscript{1}National Mobile Communications Research Laboratory,\\Frontiers Science Center for Mobile Information Communication and Security\\Southeast University, Nanjing 210096, P. R. China\\
		\textsuperscript{2}Purple Mountain Laboratories, Nanjing 211111, P. R. China\\
		\textsuperscript{3}Department of Signal Theory, Networking and Communications, University of Granada, Granada 18071\\
		\textsuperscript{4}School of Artificial Intelligence, Nanjing University of Information Science and Technology, Nanjing 210044, China\\
		\textsuperscript{5}Department of Electronic and Electrical Engineering, University College London, Torrington Place, WC1E 7JE, the UK\\
		Emails: \{zhangzhentian, dangjian, zczhang\}@seu.edu.cn, dmorales@ugr.es, c.masouros@ucl.ac.uk, jianghao@nuist.edu.cn
		}
	}

\maketitle

\begin{abstract}
The fluid antenna system (FAS) employs reconfigurable antennas for high spatial gains in compact spaces, enhancing physical layer flexibility. Channel state information (CSI) acquisition is vital for port selection and FAS optimization. Greedy algorithms rely on signal assumptions, and model-free methods face high complexity. A flexible, low-complexity solution is needed for massive connectivity in FAS. Based on expectation maximization-approximate message passing (EM-AMP) framework, efficient matrix computations and adaptive learning without prior model knowledge naturally suit CSI acquisition for FAS. We propose a EM-AMP variant exploiting FAS geographical priors, improving estimation precision, accelerating convergence, and reducing complexity in large-scale deployment. Simulations validate the efficacy of the proposed algorithm.
\end{abstract}


\section{Introduction}
\subsection{Introduction and Related Work}
Massive access, rooted in massive machine-type communication (mMTC) from IMT-2020, enables vast IoT device connectivity~\cite{mMTC1}. Fluid antenna systems (FAS)~\cite{FAS1},~\cite{FAS1.1} offer extensive physical-layer degree-of-freedom (DOF), maximizing spatial gains in compact spaces~\cite{FAS1.3},~\cite{FAS1.35},~\cite{FAS1.4}. FAS's channel response and port correlation~\cite{FAS_channel 1},~\cite{FAS_channel 5} create unique spatial diversity opportunities arising from in-depth fades and envelope response gains, enhancing massive connectivity via fluid antenna multiple access (FAMA)~\cite{FAMA1},~\cite{chernoff2}. Thus, port selection is deemed as a crucial optimization problem, which requires accurate channel state information (CSI) acquisition. Current FAS CSI acquisition estimates a subset of ports and recovers all CSI utilizing correlation among potential ports \cite{FAS1}, offering higher efficiency than full-port piloting. Specifically, CSI estimation algorithms are model-based~\cite{baseline1},~\cite{baseline2},~\cite{greedy-based},~\cite{model-based} (least squares/greedy~\cite{baseline1},~\cite{baseline2},~\cite{greedy-based} or Bayesian \cite{model-based}) or model-free \cite{model-free}, contributing to various aspects for FAS CSI acquisition.
\subsection{Challenges and Contributions}
CSI acquisition in FAS still remains challenging. Model-free methods \cite{model-free} consider the channel as a stochastic process, using kernel-based sampling and regression for distribution-free detection, but face cubic complexity with pilot length and ports, thus unsuitable for massive access. Bayesian model-based methods \cite{model-based} require impractical prior parameters. Greedy methods~\cite{baseline1},~\cite{baseline2},~\cite{greedy-based} encounter complexity and error floor issues illustrated in \cite[Fig.~12, Fig.~13]{FAS1}, \cite[Fig.~2, Fig.~4]{baseline2}. Thereby, a suitable design framework for a well balanced (low complexity yet accurate) algorithm is critically needed.

Approximate message passing (AMP) \cite{AMP1} is an iterative algorithm efficient for large matrix computations, adaptable to diverse signal models or model-free cases \cite{EM-AMP1} and widely adopted in massive access for estimation and detection ~\cite{AMP3},~\cite{AMP4},~\cite{AMP5},~\cite{EM-AMP_iotj}. Our contributions are as follows:
\begin{itemize}
	\item[-] We demonstrate that AMP-based designs are well-suited for FAS, providing stochastic signal estimation, low complexity, and robust performance. Utilizing the expectation-maximization AMP (EM-AMP) framework \cite{EM-AMP1}, we propose a novel EM-AMP exploiting geographical priors in FAS, achieving improved CSI estimation precision, faster convergence, and significantly reduced complexity with trivial performance loss.
	\item[-] Simulations show that EM-AMP greatly reduces FAS CSI estimation complexity but suffers marginal performance loss. Moreover, exploiting geographical features can promisingly resolve error floor issues reported in \cite[Fig.~12, Fig.~13]{FAS1}, \cite[Fig.~2, Fig.~4]{baseline2} with proper setups. We also investigate how antenna gap and angular information affect CSI estimation precision, suggesting promising future research directions.
\end{itemize}
\subsection{Content Structure and Notations}
Sec.~\ref{sec.2} covers system configurations and signal model; Sec.~\ref{sec.3} explains the proposed algorithm; Sec.~\ref{sec.4} presents numerical results; Sec.~\ref{sec.5} provides conclusions. {\em Notations:}~Vectors and matrices use bold lowercase and uppercase letters, respectively. Sets \(\mathbb{R}\) and \(\mathbb{C}\) denote real and complex numbers; calligraphy denotes sets, e.g., \(\mathcal{A}\). Matrix element at row \(m\), column \(n\) is \(\mathbf{A}[m,n]\). Transpose and Hermitian transpose are \((\cdot)^{\mathrm{T}}\), \((\cdot)^{\mathrm{H}}\). Complex Gaussian PDF: \(\mathcal{CN}(x;\mu,\phi)=\frac{1}{\pi\phi}e^{-\frac{|x-\mu|^2}{\phi}}\); real Gaussian PDF: \(\mathcal{N}(x;\mu,\phi)=\frac{1}{\sqrt{2\pi\phi}}e^{-\frac{(x-\mu)^2}{2\phi}}\). Modulus and \(l_2\)-norm: \(|\cdot|\), \(\|\cdot\|_2^2\).
\begin{algorithm}[htp] 
	\small
	\caption{Algorithm Baseline I, EM-AMP for FAS}
	\label{alg:algorithm1}
	\KwIn{$\mathbf{Y}$, $\mathbf{A}$, $K$, $N_o$, $G$,  $\psi$, $T_{\max}$}
	
	\textbf{Initialize:}\\
	
	$\forall k:\lambda_k^1=\frac{G}{K}\max_{a>0}\frac{1-\frac{2K}{G}[(1+a^2)\Phi(-a)-a\mathcal{N}(a;0,1)]}{1+a^2-2[(1+a^2)\Phi(-a)-a\mathcal{N}(a;0,1)]}$\hfill \text{(I1)} \\
	
	$\forall k,n:\phi^{x,1}_{k,n}=\frac{\sum_{g=1}^{G}\left|\mathbf{Y}[g,n]\right|^{2}-M\sigma_{n}^{2}}{\sum_{g=1}^{G}\sum_{k=1}^{K}\lvert \mathbf{A}[g,k]\rvert^{2}\lambda_{k}^1}$, $\mu^{x,1}_{k,n}=0$\hfill \text{(I2)} \\
	
	$\forall k,n:\tilde{x}^1_{k,n}=\int_{x}^{\infty}xp_{\mathbf{X}}\left(x;\lambda^1_k,\mu^{x,1}_{k,n},\phi^{x,1}_{k,n}\right)\mathrm{d}x$\hfill \text{(I3)} \\
	
	$\forall k,n:\tilde{\phi}^1_{k,n}=\int_{x}^{\infty}\lvert x-\tilde{x}^1_{k,n}\rvert^2 p_{\mathbf{X}}\left(x;\lambda^1_k,\mu^{x,1}_{k,n},\phi^{x,1}_{k,n}\right)\mathrm{d}x$\hfill \text{(I4)} \\
	
	$\forall g,n:\hat{s}^0_{g,n}=0$\hfill \text{(I5)}
	
	\ForEach{$t={1,2},\cdots,T_{\max}$}{
		AMP part:
		
		$\forall g,n: \hat{\phi}^{r,t}_{g,n}=\sum_{k=1}^{K}\lvert \mathbf{A}[g,k]\rvert^2 \phi^{x,t}_{k,n}$\hfill \text{(A1)}
		
		$\forall g,n: \hat{\mu}^{r,t}_{g,n}=\sum_{k=1}^{K}\mathbf{A}[g,k]\tilde{x}^t_{k,n}-\hat{\phi}^{r,t}_{g,n}\hat{s}^{t-1}_{g,n}$\hfill \text{(A2)}
		
		$\forall g,n: \tilde{\phi}^{r,t}_{g,n}=\frac{\hat{\phi}_{g,n}^{r,t}\mathbf{Y}[g,n]+\psi\hat{\mu}_{g,n}^{r,t}}{\hat{\phi}_{g,n}^{r,t}+\psi}$\hfill \text{(A3)}
		
		$\forall g,n: \tilde{\mu}^{r,t}_{g,n}=\frac{\hat{\phi}_{g,n}^{r,t}\psi}{\hat{\phi}_{g,n}^{r,t}+\psi}$\hfill \text{(A4)}
		
		$\forall g,n: \hat{\phi}^{s,t}_{g,n}=\frac{\hat{\phi}^{r,t}_{g,n}-\tilde{\phi}^{r,t}_{g,n}}{\left(\hat{\phi}^{r,t}_{g,n}\right)^2}$\hfill \text{(A5)}
		
		$\forall g,n:
		\hat{s}^t_{g,n}=\frac{\tilde{\mu}^{r,t}_{g,n}-\hat{\mu}^{r,t}_{g,n}}{\hat{\phi}^{r,t}_{g,n}}$\hfill \text{(A6)}
		
		$\forall k,n:
		\hat{\phi}^{x,t}_{k,n}=\left(\sum_{g=1}^{G}\lvert \mathbf{A}[g,k] \rvert^2\hat{\phi}^{s,t}_{g,n} \right)^{-1}$\hfill \text{(A7)}
		
		$\forall k,n:
		\hat{\mu}^{x,t}_{k,n}=\tilde{x}^t_{k,n}+\hat{\phi}^{x,t}_{k,n}\sum_{g=1}^{G}\left(\mathbf{A}[g,k]\right)^*\hat{s}^t_{g,n}$\hfill \text{(A8)}
		
		$\forall k,n:
		\gamma_{k,n}\triangleq  \frac{\hat{\mu}^{x,t}_{k,n}/\hat{\phi}^{x,t}_{k,n}+\mu^{x,t}_{k,n}/\phi^{x,t}_{k,n}}{1/\hat{\phi}^{x,t}_{k,n}+1/\phi^{x,t}_{k,n}}$\hfill \text{(B1)}
		
		$\forall k,n:
		\nu_{k,n}\triangleq \frac{1}{1/\hat{\phi}^{x,t}_{k,n}+1/\phi^{x,t}_{k,n}}$\hfill \text{(B2)}
		
		$\forall k,n:
		\beta_{k,n}\triangleq \lambda^t_k\mathcal{CN}\left(\hat{\mu}^{x,t}_{k,n};\mu^{x,t}_{k,n},\hat{\phi}^{x,t}_{k,n}+\phi^{x,t}_{k,n}\right)$\hfill \text{(B3)}
		
		$\forall k,n:
		\pi_{k,n}\triangleq\frac{1}{1+\left(\frac{\beta_{k,n}}{(1-\lambda^t_k)\mathcal{CN}(0;\hat{\mu}^{x,t}_{k,n},\hat{\phi}^{x,t}_{k,n})}\right)^{-1}}$\hfill \text{(B4)}
		
		$\forall k,n:
		\tilde{\phi}^{t+1}_{k,n}=\pi_{k,n}\left(\nu_{k,n}+|\gamma_{k,n}|^2\right)-|\pi_{k,n}\gamma_{k,n}|^2$\hfill \text{(A9)}
		
		$\forall k,n:
		\tilde{x}^{t+1}_{k,n}=\pi_{k,n}\gamma_{k,n}$\hfill \text{(A10)}
		
		EM part:
		
		$\forall k: \lambda_k^{t+1} \triangleq \frac{1}{K}\sum_{n=1}^{N_o}\pi_{k,n}$\hfill \text{(E1)}
		
		$\forall k,n: \mu^{x,t+1}_{k,n} \triangleq \frac{\sum_{k=1}^{K}\pi_{k,n}\gamma_{k,n}}{\lambda_k^{t+1}K}$\hfill \text{(E2)}
		
		$\forall k: \phi^{x,t+1}_{k,n}=\left(\tilde{x}^{t+1}_{k,n}-\mu^{x,t}_{k,n}\right)^2-\tilde{\phi}^{t+1}_{k,n}$ \hfill \text{(E3)}
		
		\uIf{$\phi^{x,t+1}_{k,n}>\phi^{x}_{\max}$}{$\phi^{x,t+1}_{k,n}=\phi^{x}_{\max}$}\ElseIf{$\phi^{x,t+1}_{k,n}<\phi^{x}_{\min}$}{$\phi^{x,t+1}_{k,n}=\phi^{x}_{\min}$}
	}
	\KwOut{$\lambda_{k},k\in\left\{1,\ldots,K\right\}$, $\tilde{\mathbf{X}}[k,n]=\tilde{x}^{t+1}_{k,n},k\in\left\{1,\ldots,K\right\}, n\in\left\{1,\ldots,N_o\right\}$.}
\end{algorithm}
\section{System Descriptions}\label{sec.2}
\subsection{System Configuration}\label{System Configuration}
Consider an uplink scenario where a base station (BS) with a $W$-length fluid antenna serves $K$ single-antenna users in service area with radius $d_{ref}\le d_k\le d_{\max}$. Transmission is organized into frames, each with $G$ pilot symbols for channel training. Traffic is sporadic, with $K_a$ active users per frame. Each user has a unique pilot signature, $\mathbf{a}_k\sim \mathcal{CN}\left(0,1/G\right) \in \mathbb{C}^{G \times 1}$, and the BS restores a pilot codebook $\mathbf{A} = [\mathbf{a}_1, \ldots, \mathbf{a}_K] \in \mathbb{C}^{G \times K}$, where $\|\mathbf{a}_k\|_2^2 = 1$. Due to limited hardware overhead, only $N_o$ equi-spaced ports with gap width $\frac{W}{N_o-1}$ can be potentially activated at the BS. The channel coefficient between the $k$-th user and the BS is $\mathbf{h}_k \in \mathbb{C}^{1 \times N_o}$.

Omitting asynchronous transmission, the received signal at the BS is:
\begin{equation}\label{eq:1}
	\mathbf{Y} = \sum_{k=1}^{K} \alpha_k \mathbf{a}_k \mathbf{h}_k + \mathbf{Z},
\end{equation}
where $\mathbf{Y} \in \mathbb{C}^{G \times N_o}$ is the received signal, $\alpha_k$ indicates whether the $k$-th pilot is active ($\alpha_k = 1$) or idle ($\alpha_k = 0$), and $\mathbf{Z}$ is the i.i.d. AWGN with zero mean and variance $\psi$, i.e., $\mathcal{CN}(0, \psi)$. The compact form of \eqref{eq:1} is:
\begin{equation}\label{eq:2}
	\mathbf{Y} = \mathbf{A} \mathbf{X} + \mathbf{Z},
\end{equation}
where $\mathbf{A}$ is the pilot codebook and $\mathbf{X} \in \mathbb{C}^{K \times N_o}$ is a row-sparse matrix with only $K_a$ non-zero rows, representing a compressive sensing model.
\subsection{FAS Channel Model}\label{FAS Channel Model}
The channel vector $\mathbf{h}_k$ consists of small-scale fading coefficient (LSFC) $\mathbf{s}_k$ and large-scale fading $\varsigma_k$, i.e., $\mathbf{h}_k = \sqrt{\varsigma_k} \mathbf{s}_k$. For small-scale fading, we use a geometric model with $L_s$ scattering paths. Let $\sigma_{k,l}$ and $\theta_{k,l}$ represent the path strength and angle-of-arrival (AoA) of the $k$-th user at the $l$-th path. The receiving antenna is a linear array with length $W = \frac{\lambda_{len}}{2}(M-1)$, and $N_o$ ports are uniformly spaced with gap width $\frac{W}{N_o-1}$. The normalized steering response for the $l$-th path is:
\begin{equation}
	\mathbf{s}_{k,l} = \frac{\exp\left(-j\frac{2\pi(n-1)W}{(N_o-1)\lambda_{len}} \cos \theta_{k,l}\right)}{\sqrt{N_o}}, \quad n \in \{1, \ldots, N_o\}.
\end{equation}
Thus, small-scale fading is:
\begin{equation}\label{eq:4}
	\mathbf{s}_k = \sum_{l=1}^{L_s} \sigma_{k,l} \mathbf{s}_{k,l} \in \mathbb{C}^{1 \times N_o}.
\end{equation}
Large-scale fading is determined by the distance $d_k$ between the $k$-th user and the BS via a function $\varsigma_k = f(d_k)$. Small-scale fading is normalized such that $\mathbb{E}\left\{\|\mathbf{s}_k\|_2^2\right\} = N_o$, and thus $\mathbb{E}\left\{\|\mathbf{h}_k\|_2^2\right\} = N_o \varsigma_k$. The scattering path model \eqref{eq:4} is categorized by non-line-of-sight (NLOS) or mixed NLOS/LOS components, impacting \(\sigma_{k,l}\). NLOS-only paths, due to dispersive obstacles, lack direct transmission signals. For mixed LOS/NLOS, with Rician factor \(K_r\), the path strength is \(\sqrt{\frac{K_r\Omega}{K_r+1}}e^{j\beta_k}\), where \(\beta_k\) is the LOS phase and \(\Omega\) is a scaling constant. The remaining \(L_s-1\) NLOS path amplitudes satisfy \(\sqrt{\sum_{l=1}^{L_s-1} \sigma_{k,l}^2} = \sqrt{\frac{\Omega}{K_r+1}}\), with LOS AoA having larger path strength than NLOS AoAs.
\section{Proposed Algorithm}\label{sec.3}
\subsection{Proposed EM-AMP Exploiting Geographical Diversity}
Following the EM-AMP framework \cite{EM-AMP1}, we introduce the proposed update rule exploiting geographical information. The priori distribution of \(\mathbf{X}\) in \eqref{eq:2} follows a Bernoulli-Gaussian (BG) model:
\begin{equation}\label{eq:10}
	\begin{aligned}
			&p_{\mathbf{X}}(x_{k,n}; \lambda_k, \mu^{x}_{k,n}, \phi^{x}_{k,n}) \\
			&= (1-\lambda_k)\delta(x_{k,n}) + \lambda_k \mathcal{CN}(x_{k,n}; \mu^{x}_{k,n}, \phi^{x}_{k,n}),
	\end{aligned}
\end{equation}
where \(\lambda_k\) is the activity probability of the \(k\)-th codeword, \(\mu^{x}_{k,n}\) and \(\phi^{x}_{k,n}\) are the mean and variance of the signal, and \(\mathbf{q}_k = (\lambda_k, \mu^{x}_{k,n}, \phi^{x}_{k,n}, \psi)\) aggregates prior parameters estimated from noisy observations. Posterior estimates are denoted with a hat, e.g., \(\hat{a}\). AMP models the noisy output \(y_{g,n}\) and noise-free output \(r_{g,n} = \mathbf{a}_g^{\mathrm{T}} \mathbf{x}_n\) (noise-free matrix is denoted by $\mathbf{R}$), where \(\mathbf{a}_g^{\mathrm{T}}\) is the \(g\)-th row of \(\mathbf{A}\), \(\mathbf{x}_n\) is the \(n\)-th column of \(\mathbf{X}\), \(g \in \{1, \ldots, G\}\), and \(n \in \{1, \ldots, N_o\}\). The conditional PDF is:
\begin{equation}\label{eq:11}
	p_{\mathbf{Y}|\mathbf{R}}(y_{g,n} | r_{g,n}; \mathbf{q}) = \mathcal{CN}(y_{g,n}; r_{g,n}, \psi).
\end{equation}
The marginal posterior of the noise-free output is:
\begin{equation}\label{eq:12}
	\small
	\begin{aligned}
		&p_{\mathbf{R}|\mathbf{Y}}(r_{g,n} | \mathbf{y}_n; \hat{\mu}_{g,n}^r, \hat{\phi}_{g,n}^r, \mathbf{q}) \\
		&\triangleq \frac{p_{\mathbf{Y}|\mathbf{R}}(y_{g,n} | r_{g,n}; \mathbf{q}) \mathcal{CN}(r_{g,n}; \hat{\mu}_{g,n}^r, \hat{\phi}_{g,n}^r)}{\int_r p_{\mathbf{Y}|\mathbf{R}}(y_{g,n} | r; \mathbf{q}) \mathcal{CN}(r; \hat{\mu}_{g,n}^r, \hat{\phi}_{g,n}^r)},
	\end{aligned}
\end{equation}
where \(\hat{\mu}_{g,n}^r\) and \(\hat{\phi}_{g,n}^r\) are iteration-dependent \cite[Table~I, R2-R1]{EM-AMP1} and computed via (A1)-(A2) in Algorithm~\ref{alg:algorithm1}. Using Gaussian identities ($\mathrm{E}\left\{\mathcal{CN}(x;a,A)\mathcal{CN}(x;b,B)\right\} =\frac{aB+bA}{A+B},\mathrm{var}\left\{\mathcal{CN}(x;a,A)\mathcal{CN}(x;b,B)\right\} =\frac{AB}{A+B}$), the posterior statistics are:
\begin{subequations}\label{eq:13}
	\small
	\begin{align}
		\mathrm{E}_{\mathbf{R}|\mathbf{Y}}(r_{g,n} | \mathbf{y}_n; \hat{\mu}_{g,n}^r, \hat{\phi}_{g,n}^r, \mathbf{q}) &= \frac{\hat{\phi}_{g,n}^r \mathbf{Y}[g,n] + \psi \hat{\mu}_{g,n}^r}{\hat{\phi}_{g,n}^r + \psi}, \label{eq:13a}\\
		\mathrm{var}_{\mathbf{R}|\mathbf{Y}}(r_{g,n} | \mathbf{y}_n; \hat{\mu}_{g,n}^r, \hat{\phi}_{g,n}^r, \mathbf{q}) &= \frac{\hat{\phi}_{g,n}^r \psi}{\hat{\phi}_{g,n}^r + \psi},\label{eq:13b}
	\end{align}
\end{subequations}
denoted as \(\tilde{\mu}_{g,n}^r\) and \(\tilde{\phi}_{g,n}^r\). AMP approximates the marginal posterior of \(\mathbf{X}\):
\begin{equation}\label{eq:14}
	\begin{aligned}
	&p_{\mathbf{X}|\mathbf{Y}}(x_{k,n}|\mathbf{y}_n;\hat{\mu}^{x}_{k,n},\hat{\phi}^{x}_{k,n},\mathbf{q}_k) \\
	&\triangleq \frac{p_\mathbf{x}(x_{k,n};\mathbf{q}_k)\mathcal{CN}(x_{k,n};\hat{\mu}^{x}_{k,n},\hat{\phi}^{x}_{k,n})}{\underbrace{\int_{x} p_\mathbf{x}(x;\mathbf{q}_k)\mathcal{CN}(x;\hat{\mu}^{x}_{k,n},\hat{\phi}^{x}_{k,n})}_{\zeta_{k,n}}},
	\end{aligned}
\end{equation}
where \(\hat{\mu}^{x}_{k,n}\), \(\hat{\phi}^{x}_{k,n}\) are computed via (A7)-(A8) in Algorithm~\ref{alg:algorithm1}, and:
\begin{subequations}\label{eq:15}
	\begin{align}
		\zeta_{k,n}&=\int_{x} p_\mathbf{x}(x;\mathbf{q}_k)\mathcal{CN}(x;\hat{\mu}^{x}_{k,n},\hat{\phi}^{x}_{k,n})\label{eq:15a}\\
		&=(1-\lambda_k)\mathcal{CN}(0;\hat{\mu}^x_{k,n},\hat{\phi}^x_{k.n})+\label{eq:15b}\\ &\lambda_k\mathcal{CN}(0;\hat{\mu}^x_{k,n}-\mu^x_{k,n},\hat{\phi}^x_{k,n}+\phi^x_{k,n}) \nonumber
	\end{align}
\end{subequations}
Substituting \eqref{eq:10} into \eqref{eq:15a}, the posterior distribution is expressed as a BG model:
\begin{equation}\label{eq:16}
	\begin{aligned}
		&p_{\mathbf{X}|\mathbf{Y}}(x_{k,n} | \mathbf{y}_n; \hat{\mu}^{x}_{k,n}, \hat{\phi}^{x}_{k,n}, \mathbf{q}_k)\\
		& \triangleq (1-\pi_{k,n}) \delta(x_{k,n}) + \pi_{k,n} \mathcal{CN}(x_{k,n}; \gamma_{k,n}, \nu_{k,n}),
	\end{aligned}
\end{equation}
with parameters:
\begin{subequations}\label{eq:17}
	\begin{align}
		\gamma_{k,n} &\triangleq \frac{\hat{\mu}_{k,n}^x / \hat{\phi}_{k,n}^x + \mu_{k,n}^x / \phi_{k,n}^x}{1 / \hat{\phi}_{k,n}^x + 1 / \phi_{k,n}^x}, \\
		\nu_{k,n} &\triangleq \frac{1}{1 / \hat{\phi}_{k,n}^x + 1 / \phi_{k,n}^x}, \\
		\beta_{k,n} &\triangleq \lambda_k \mathcal{CN}(\hat{\mu}_{k,n}^x; \mu_{k,n}^x, \hat{\phi}_{k,n}^x + \phi_{k,n}^x), \\
		\pi_{k,n} &\triangleq \frac{1}{1 + \frac{(1-\lambda_k) \mathcal{CN}(0; \hat{\mu}_{k,n}^x, \hat{\phi}_{k,n}^x)}{\beta_{k,n}}},
	\end{align}
\end{subequations}
where \(\pi_{k,n} \in [0,1]\) is the likelihood of \(x_{k,n} \neq 0\). The activity probability is \(\lambda_k = \frac{1}{N_o} \sum_{n=1}^{N_o} \pi_{k,n}\). Posterior statistics are:
\begin{subequations}\label{eq:18}
	\begin{align}
		&\mathrm{E}_{\mathbf{X}|\mathbf{Y}}(x_{k,n} | \mathbf{y}_n; \hat{\mu}_{k,n}^x, \hat{\phi}_{k,n}^x, \mathbf{q}_k) = \pi_{k,n} \gamma_{k,n}, \label{eq:18a}\\
		&\mathrm{var}_{\mathbf{X}|\mathbf{Y}}(x_{k,n} | \mathbf{y}_n; \hat{\mu}_{k,n}^x, \hat{\phi}_{k,n}^x, \mathbf{q}_k) \nonumber \\
		&= \pi_{k,n} (\nu_{k,n} + |\gamma_{k,n}|^2) - |\pi_{k,n} \gamma_{k,n}|^2,\label{eq:18b}
	\end{align}
\end{subequations}
denoted as \(\tilde{x}_{k,n}\) and \(\tilde{\phi}_{k,n}^x\) respectively. The AMP calculations require \(\mathbf{q}_k\), learned iteratively, forming the E-step \cite[Eq.18-Eq.21]{EM-AMP1}. The M-step is:
\begin{equation}\label{eq:19}
	\mathbf{q}_k^{t+1} = \arg\max_{\mathbf{q}_k^t} \hat{\mathrm{E}} \{ \ln p_{\mathbf{X}}(\mathbf{X}; \mathbf{q}_k) \mid \mathbf{Y}; \mathbf{q}_k^t \},
\end{equation}
where \(\hat{\mathrm{E}}\) uses AMP’s posterior approximation. Prior parameters update as:
\begin{subequations}\label{eq:20}
	\begin{align}
		\lambda_k^{t+1} &= \frac{1}{K} \sum_{n=1}^{N_o} \pi_{k,n},~ 
		\mu_{k,n}^{x,t+1} = \frac{\sum_{k=1}^{K} \pi_{k,n} \gamma_{k,n}}{\lambda_k^{t+1} K}, \\
\phi^{x,t+1}_{k,n}&=\left\{\begin{matrix}
	\phi^x_{\min},& \mathrm{if}~\frac{\sum_{n=1}^{N_o}V_{k,n}}{\sum_{n=1}^{N_o}\pi_{k,n}}<\phi^x_{\min} \\ 
	\frac{\sum_{n=1}^{N_o}V_{k,n}}{\sum_{n=1}^{N_o}\pi_{k,n}}, & \mathrm{if}~\phi^x_{\min}\le\frac{\sum_{n=1}^{N_o}V_{k,n}}{\sum_{n=1}^{N_o}\pi_{k,n}}\le \phi^x_{\max}\\
	\phi^x_{\max}.& \mathrm{if}~\frac{\sum_{n=1}^{N_o}V_{k,n}}{\sum_{n=1}^{N_o}\pi_{k,n}}>\phi^x_{\max}
\end{matrix}\right.
	\end{align}
\end{subequations}
\subsection{Derivations: Variance $\phi_{k,n}^{x}$ Update Rule}
The geographical information, specifically the LSFC \(\varsigma_k\) from the channel model in Section~\ref{FAS Channel Model}, determines the variance \(\phi^x_{k,n}\) of the prior distribution \(p_{\mathbf{X}}(x_{k,n}; \lambda_k, \mu^{x}_{k,n}, \phi^{x}_{k,n})\) in \eqref{eq:10}. Since \(\phi^x_{k,n} = f(d_k)\) correlates with distance on a 2-D plane \cite{LSFC1,LSFC2,LSFC3}, this geographical information can be used to update \(\phi^x_{k,n}\), denoted as \(\phi^x_{k,n}(d_k)\). Following the EM principle and incremental updating \cite{EM-AMP1,EM1,EM2}, the distance \(d_k\) for each user can be sequentially estimated, similar to \eqref{eq:19}, leveraging the independence of users' locations.
\begin{equation}\label{eq:21}
	\small
	\begin{aligned}
		&d_k^{t+1} =\mathop{\arg\max}_{d_{ref}\le d_k\le d_{\max}}\sum_{n=1}^{N_o} \hat{\mathrm{E}}\left\{\ln p_{\mathbf{X}}\left(x_{k,n};\mathbf{q}_k\right)|\mathbf{Y},\mathbf{q}^t_k\right\}\\
		&=\mathop{\arg\max}_{d_{ref}\le d_k\le d_{\max}}\sum_{n=1}^{N_o}\underbrace{\int_{x_{k,n}}p_{\mathbf{X}|\mathbf{Y}}(x_{k,n}|\mathbf{y}_n;\mathbf{q}^t_k)\ln p_{\mathbf{X}}\left(x_{k,n};\mathbf{q}^t_k\right)}_{\triangleq J\left(\phi_{k,n}\right)},
	\end{aligned}
\end{equation}
where posterior PDF $p_{\mathbf{X}|\mathbf{Y}}(x_{k,n}|\mathbf{y}_n;\mathbf{q}^t_k)$ and prior PDF $p_{\mathbf{X}}\left(x_{k,n};\mathbf{q}^t_k\right)$ are identical to \eqref{eq:16} and \eqref{eq:10} respectively and we denote the integral in \eqref{eq:21} by $J\left(\phi_{k,n}\right)$ in the sequel.

Meanwhile, the integral area should be split into separate domains considering that the logarithmic term in $J\left(\phi^x_{k,n}\right)$ has different expressions:
\begin{equation}\label{eq:22}
	p_{\mathbf{X}}\left(x_{k,n};\mathbf{q}^t_k\right)=\left\{\begin{matrix}
		(1-\lambda^t_k)\delta\left(x_{k,n}\right),& x_{k,n}=0\\
		\lambda^t_k\mathcal{CN}\left(x_{k,n};\mu^{x,t}_{k,n},\phi^{x,t}_{k,n}\left(d_k\right)\right),& x_{k,n}\neq 0.
	\end{matrix}\right.
\end{equation}
\begin{figure*}[!t]
	\normalsize
	\begin{subequations}\label{eq:23}
			\small
		\begin{align}
			&J\left(\phi^x_{k,n}\right) = \lim_{\epsilon \to 0}\int_{x_{k,n}\in \mathcal{B}_{\epsilon}} p_{\mathbf{X}|\mathbf{Y}}(x_{k,n}|\mathbf{y}_n;\mathbf{q}^t_k)\ln p_{\mathbf{X}}\left(x_{k,n};\mathbf{q}^t_k\right) 		
			+		
			\lim_{\epsilon \to 0}\int_{x_{k,n}\in \overline{\mathcal{B}}_{\epsilon}} p_{\mathbf{X}|\mathbf{Y}}(x_{k,n}|\mathbf{y}_n;\mathbf{q}^t_k)
			\ln p_{\mathbf{X}}\left(x_{k,n};\mathbf{q}^t_k\right) \nonumber\\
			&= \underbrace{\lim_{\epsilon \to 0}\int_{x_{k,n}\in \mathcal{B}_{\epsilon}} p_{\mathbf{X}|\mathbf{Y}}(x_{k,n}|\mathbf{y}_n;\mathbf{q}^t_k)
				\ln \left[(1-\lambda^t_k)\delta\left(x_{k,n}\right)\right]}_{\triangleq C_{k,n}} 
			+	
			\lim_{\epsilon \to 0}\int_{x_{k,n}\in \overline{\mathcal{B}}_{\epsilon}} p_{\mathbf{X}|\mathbf{Y}}(x_{k,n}|\mathbf{y}_n;\mathbf{q}^t_k) 
			\ln \left[\lambda^t_k\mathcal{CN}\left(x_{k,n};\mu^{x,t}_{k,n},\phi^{x,t}_{k,n}\left(d_k\right)\right)\right] \label{eq:23a}\\
			&=C_{k,n}
			+
			\lim_{\epsilon \to 0}\int_{x_{k,n}\in \overline{\mathcal{B}}_{\epsilon}} p_{\mathbf{X}|\mathbf{Y}}(x_{k,n}|\mathbf{y}_n;\mathbf{q}^t_k) 
			\ln \left[
			\frac{\lambda^t_k}{\pi \phi^{x,t}_{k,n}\left(d_k\right)}\exp\left\{-\frac{\lvert x_{k,n}-\mu^{x,t}_{k,n}\rvert^2}{\phi^{x,t}_{k,n}\left(d_k\right)}
			\right\}
			\right]\nonumber\\
			&=C_{k,n}
			+\ln \left(
			\frac{\lambda^t_k}{\pi \phi^{x,t}_{k,n}\left(d_k\right)}\right)
			\underbrace{ \lim_{\epsilon \to 0}\int_{x_{k,n}\in \overline{\mathcal{B}}_{\epsilon}} p_{\mathbf{X}|\mathbf{Y}}(x_{k,n}|\mathbf{y}_n;\mathbf{q}^t_k)}_{\triangleq \pi_{k,n}}
			-\frac{1}{\phi^{x,t}_{k,n}\left(d_k\right)}\underbrace{\lim_{\epsilon \to 0}\int_{x_{k,n}\in \overline{\mathcal{B}}_{\epsilon}} p_{\mathbf{X}|\mathbf{Y}}(x_{k,n}|\mathbf{y}_n;\mathbf{q}^t_k) \lvert x_{k,n}-\mu^{x,t}_{k,n}\rvert^2}_{\triangleq V_{k,n}}\label{eq:23b}\\
			&=C_{k,n}+\pi_{k,n}\ln \left(
			\frac{\lambda^t_k}{\pi \phi^{x,t}_{k,n}\left(d_k\right)}\right)-\frac{V_{k,n}}{\phi^{x,t}_{k,n}\left(d_k\right)}.\label{eq:23c}
		\end{align}
	\end{subequations}
	\hrulefill
\end{figure*}
Accordingly, the integral area is split into two parts denoted by $\mathcal{B}_{\epsilon}=\left[-\epsilon,\epsilon\right]$ and $\overline{\mathcal{B}}_{\epsilon}=\mathbb{C}\setminus \mathcal{B}_{\epsilon}$, where $\epsilon\to 0$ controls the borders between $\mathcal{B}_{\epsilon}$ and $\overline{\mathcal{B}}_{\epsilon}$. The integral process is given in \eqref{eq:23} resulting in $J\left(\phi^x_{k,n}\right)=C_{k,n}+\pi_{k,n}\ln \left(
\frac{\lambda^t_k}{\pi \phi^{x,t}_{k,n}\left(d_k\right)}\right)-\frac{V_{k,n}}{\phi^{x,t}_{k,n}\left(d_k\right)}$. In \eqref{eq:23a}, $C_{k,n}$ is shown as a constant irrelevant to $\phi^x_{k,n}$. For \eqref{eq:23b}, two major integral components ($ \pi_{k,n}$ and $V_{k,n}$) are calculated as:
\begin{subequations}\label{eq:24}
	\begin{align}
		&\lim_{\epsilon \to 0}\int_{x_{k,n}\in \overline{\mathcal{B}}_{\epsilon}} p_{\mathbf{X}|\mathbf{Y}}(x_{k,n}|\mathbf{y}_n;\mathbf{q}^t_k) \nonumber\\
		&=\lim_{\epsilon \to 0}\int_{x_{k,n}\in \overline{\mathcal{B}}_{\epsilon}}\pi_{k,n}\mathcal{CN}\left(x_{k,n};\gamma_{k,n},\nu_{k,n}\right)=\pi_{k,n},\label{eq:24a}\\
		&V_{k,n}=\lim_{\epsilon \to 0}\int_{x_{k,n}\in \overline{\mathcal{B}}_{\epsilon}} p_{\mathbf{X}|\mathbf{Y}}(x_{k,n}|\mathbf{y}_n;\mathbf{q}^t_k) \lvert x_{k,n}-\mu^{x,t}_{k,n}\rvert^2 \nonumber \\
		&=\left[\mathrm{E}_{\mathbf{X}|\mathbf{Y}}\left(x_{k,n}|\mathbf{y}_n;\mathbf{q}^t_k\right)-\mu^{x,t}_{k,n}\right]^2-\mathrm{Var}_{\mathbf{X}|\mathbf{Y}}\left(x_{k,n}|\mathbf{y}_n;\mathbf{q}^t_k\right)\nonumber \\
		&=\left(\underbrace{\pi_{k,n}\gamma_{k,n}}_{\tilde{x}^{t+1}_{k,n}}-\mu^{x,t}_{k,n}\right)^2-\underbrace{\pi_{k,n}\left(\nu_{k,n}+|\gamma_{k,n}|^2\right)+|\pi_{k,n}\gamma_{k,n}|^2}_{\tilde{\phi}^{t+1}_{k,n}},\label{eq:24b}
	\end{align}
\end{subequations}
where $\mathrm{E}_{\mathbf{X}|\mathbf{Y}}\left(x_{k,n}|\mathbf{y}_n;\mathbf{q}^t_k\right)$ and $\mathrm{Var}_{\mathbf{X}|\mathbf{Y}}\left(x_{k,n}|\mathbf{y}_n;\mathbf{q}^t_k\right)$ are identical to the statistics in \eqref{eq:18} omitting irrelevant terms and have been calculated before EM update during (A9)-(A10) in Algorithm~\ref{alg:algorithm1}. Therefore, the EM maximization expression in \eqref{eq:21} is converted into:
\begin{equation}\label{eq:25}
		\small
	\begin{aligned}
		&d_k^{t+1}=f^{-1}\left(\phi^{x,t+1}_{k,n}\right)\\
		&=\mathop{\arg\max}_{\phi^{x,t}_{k,n}}\sum_{n=1}^{N_o}C_{k,n}+\pi_{k,n}\ln \left(
		\frac{\lambda^t_k}{\pi \phi^{x,t}_{k,n}\left(d_k\right)}\right)-\frac{V_{k,n}}{\phi^{x,t}_{k,n}\left(d_k\right)}\\
		&\Rightarrow \mathop{\arg\max}_{\phi^{x,t}_{k,n}}\sum_{n=1}^{N_o}
		\pi_{k,n}\ln \left(
		\frac{\lambda^t_k}{\pi \phi^{x,t}_{k,n}\left(d_k\right)}\right)-\frac{V_{k,n}}{\phi^{x,t}_{k,n}\left(d_k\right)}\\
		&=\mathop{\arg\min}_{\phi^{x,t}_{k,n}}\sum_{n=1}^{N_o}
		\pi_{k,n}\ln \left(
		\frac{\pi}{\lambda^t_k}\right)+\pi_{k,n}\ln \left(\phi^{x,t}_{k,n}\left(d_k\right)\right)+\frac{V_{k,n}}{\phi^{x,t}_{k,n}\left(d_k\right)}\\
		&\Rightarrow \mathop{\arg\min}_{\phi^{x,t}_{k,n}}\sum_{n=1}^{N_o}\pi_{k,n}\ln \left(\phi^{x,t}_{k,n}\left(d_k\right)\right)+\frac{V_{k,n}}{\phi^{x,t}_{k,n}\left(d_k\right)},
	\end{aligned}
\end{equation}
where components $C_{k,n}$ and $\pi_{k,n}\ln \left(\phi^{x,t}_{k,n}\left(d_k\right)\right)$ are omitted since they are irrelevant to $\phi^{x,t}_{k,n}$. Since the variance contributed by LSFC should be identical among all receiving antennas, one can set the first-derivative of \eqref{eq:25} to zero and find a closed-form solution to update the prior PDF variance:
\begin{equation}\label{eq:26}
	\phi^{x,t+1}_{k,n}=\frac{\sum_{n=1}^{N_o}V_{k,n}}{\sum_{n=1}^{N_o}\pi_{k,n}},
\end{equation}
where the intermediate parameter $V_{k,n}$ is calculated in \eqref{eq:24b}, and since $\phi^{x,t+1}_{k,n}$ is assumed to be correlated with geographical prior in 2-D domain by function $f\left(d_k\right),~ d_{ref}\le d_k \le d_{\max}$, the function presumably has minimum ($\phi^x_{\min}$) and maximum ($\phi^x_{\max}$) values, i.e., $\phi^x_{\min}\le\phi^{x,t+1}_{k,n}\le \phi^x_{\max}$.
\subsection{Complexity Analyses}
The proposed algorithm's complexity is mainly determined by steps (A1)-(A2) and (A7)-(A8) in Algorithm~\ref{alg:algorithm1}, involving matrix multiplications with complexity \(\mathcal{O}(4KGN_o)\) per iteration. The EM component updates \(\lambda_k\) with complexity \(\mathcal{O}(KN_o)\), and the prior PDF mean and variance with \(\mathcal{O}(2KN_o+2K)\). Total complexity is \(\mathcal{O}(4KGN_o + 3KN_o + 2K)\), which is independent of \(K_a\) and thus suitable for massive connectivity.
\begin{table}[htp]
	\caption{System Configurations}
	\label{tab:system configurations}
	\centering
	\scalebox{0.87}{
		\begin{threeparttable}
	\begin{tabular}{lll}
		\hline
		Parameter       & Definitions               & Setups      \\ \hline
		$d_{\max}$      & Far field upper range     & 500 meters  \\
		$d_{def}$       & Far field lower range     & 50 meters   \\
		$\theta_{\max}$ & FAS AoA angle upper range & 150 degrees \\
		$\theta_{\min}$ & FAS AoA angle upper range & 30 degrees  \\
		$f(d_k)$        & LSFC function             & $d_k^{-2}$  \\
		$L_s$           & Scattering path num       & 3           \\ 
		$K_r$           & Rician factor				& 2\\
		$T_{\max}$      & AMP iteration upper range & 15\\
		$K$ 			& Total user num			& 1000\\
		$G$				& Pilot length				& 400\\
		$N_s$           & AoA sample num			& 121 (resolution $1^o$) \\
		$M$             & Antenna length constant			& 64, $W = \frac{\lambda_{len}(M-1)}{2}$\\
		\hline
	\end{tabular}
	\begin{tablenotes}
		\item The simulator configurations are 13th Gen Intel(R) Core(TM) i7-13700 (2.10 GHz), 32.0 GB RAM, Windows 11-24H2 with MATLAB R2024b.
	\end{tablenotes}
	\end{threeparttable}
}
\end{table}
\section{Numerical Results}\label{sec.4}
\subsection{Parameter Setups and Performance Metric}
In Table.~\ref{tab:system configurations}, \textbf{universal parameter} setups are summarized, which remains unchanged unless stated otherwise. \textbf{The performance metrics} include activity detection error (ADE) and channel estimation normalized mean square error (NMSE):
\begin{equation}
	\small
		\mathrm{ADE}= 1-\frac{\lvert \mathcal{A}\cap \tilde{\mathcal{A}} \rvert}{K_a},~ 
		\mathrm{NMSE}= \frac{\mathrm{E}\left\{\|\mathbf{h}_k-\tilde{\mathbf{h}}_k\|_2^2\right\}}{\mathrm{E}\left\{\|\mathbf{h}_k\|_2^2\right\}}
\end{equation}
where $\mathcal{A},\mathbf{h}_k,$ denote the true activity set, channel coefficients prior, and $\tilde{\mathcal{A}},\tilde{\mathbf{h}}_k$ are the corresponding estimated quantities. Only NMSE of correctly detected users will be averaged. Moreover, the \textit{received} signal-to-noise ratio (SNR) is defined as $\mathrm{SNR}=\frac{\|\mathbf{a}_k\|_2^2\mathrm{E}\left\{\|\mathbf{h}_k\|_2^2\right\}}{\mathrm{E}\left\{\|\mathbf{Z}\|_{\mathsf{F}}^2\right\}}=\frac{N_o\bar{\varsigma}_k}{\psi GN_o}=\frac{\bar{\varsigma}_k}{G\psi}$, where $\bar{\varsigma}_k=\frac{1}{K_a}\sum_{k=1}^{K_a}\varsigma_k$ is the averaged LSFC.

\textbf{The baseline algorithms} designed for FAS are AoA \textbf{codebook-based} \cite{baseline1} and \textbf{least squares} \cite{baseline2}, both using simultaneous matching pursuit (SOMP) \cite{ODMA_iotj} for activity detection. SOMP leverages multiple measurements from \(N_o\) ports, significantly outperforming OMP used in \cite{baseline2}. Moreover, the proposed algorithm, tailored for FAS, is also compared with \textbf{conventional} EM-AMP \cite{EM-AMP1,AMP3} to highlight its superior performance within certain aspects.
\subsection{Performance Evaluation}
\begin{figure}[!t]
	\centering
	\includegraphics[width=3.2in]{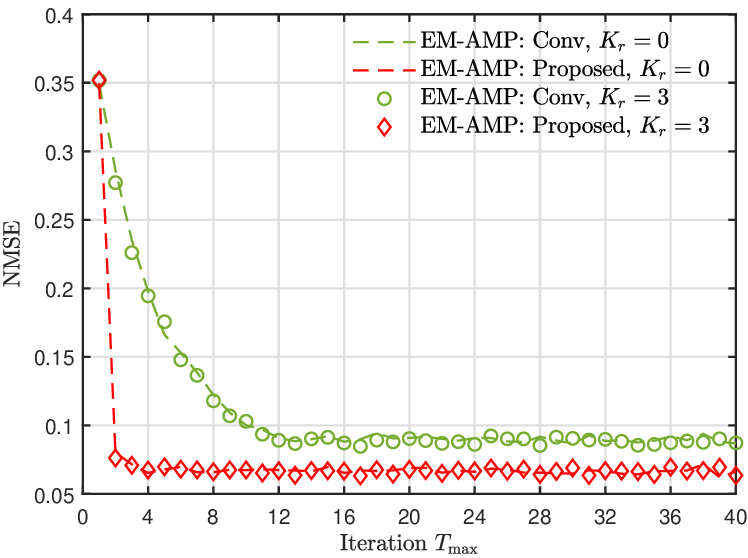}
	\caption{Convergence behavior of the proposed scheme under NLOS-only (\(K_r=0\)) and NLOS/LOS (\(K_r=3\)) with \(N_o=8\) active ports, $K_a=10$ users and $\mathrm{SNR}=-14$ dB.}\label{sim:convergence}
\end{figure}
\subsubsection{Convergence Behavior}
Fig.~\ref{sim:convergence} examines EM-AMP convergence with \(N_o=8\) active ports, 10 active users, and -14 dB SNR. Incorporating geographical priors in FAS significantly accelerates convergence by constraining the prior PDF variance search span. And all can function well under NLOS-only \(K_r=0\) or NLOS/LOS \(K_r=3\).
\begin{figure}[!t]
	\centering
	\includegraphics[width=3in]{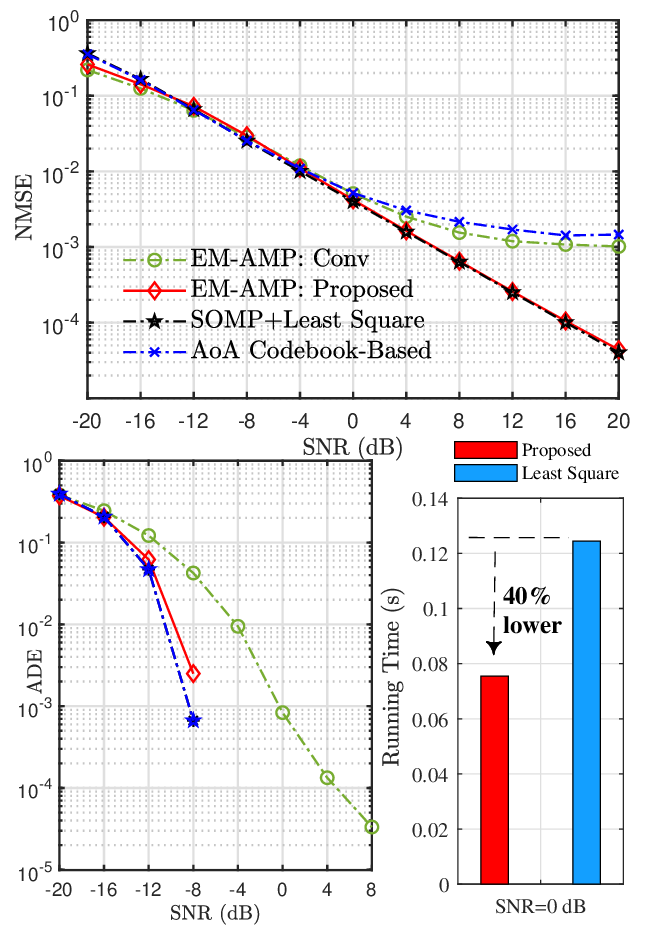}
	\caption{Illustration of ADE, NMSE and running time (s) of different algorithms versus SNR (dB) with $N_o=8$ active ports and $K_a=150$ active users. Performance baselines in comparison are conventional EM-AMP \cite{EM-AMP1,AMP3}, SOMP+Least Square \cite{baseline2} and AoA codebook-based \cite{baseline1}.}\label{sim:Performance_SNR}
\end{figure}
\subsubsection{Performance versus SNR (dB)}
Fig.~\ref{sim:Performance_SNR} shows algorithm performance and complexity comparison with \(N_o=8\), 150 active users across varying SNR. The proposed EM-AMP, leveraging geographical features, outperforms least squares and AoA codebook-based methods below \(-16\) dB SNR, with similar NMSE and ADE thereafter. More importantly, it reduces computational complexity by nearly 40\%, highlighting its superiority and favorable effective precision-complexity trade-off.

Notably, NMSE curves of conventional EM-AMP and AoA codebook-based methods converge similarly above 4 dB SNR, consistent with \cite[Fig.~12, Fig.~13]{FAS1}, \cite[Fig.~2, Fig.~4]{baseline2}, where only least squares reduces channel estimation NMSE with increasing SNR. The cause of this estimation floor remains unclear. Once again, it demonstrates the importance of exploiting geographical feature for FAS CSI acquisition.
\begin{figure}[!t]
	\centering
	\includegraphics[width=3.5in]{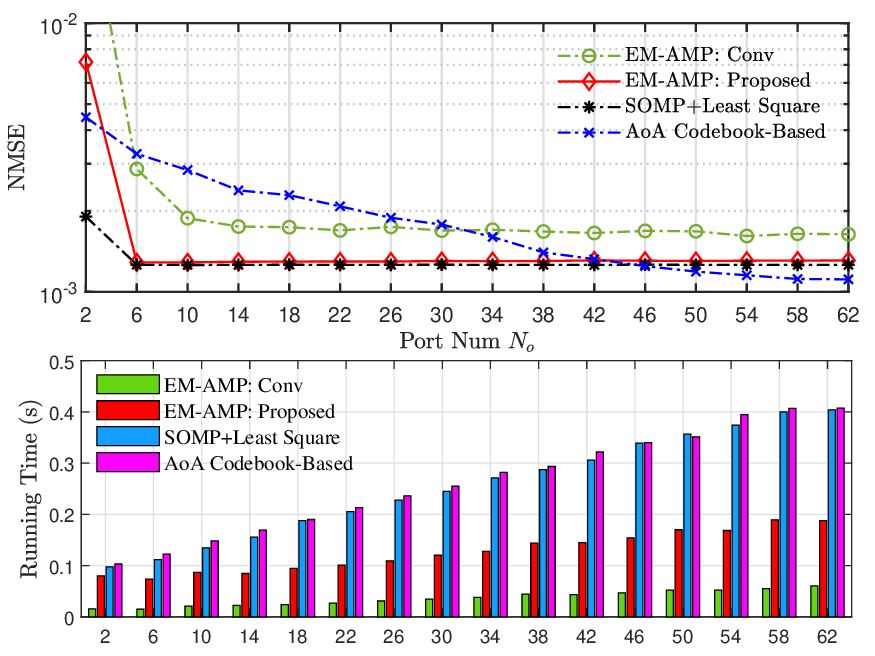}
	\caption{Illustration of ADE, NMSE  and running time (s) of different algorithms versus number of active ports $N_o$ with $\mathrm{SNR}=5$ dB and $K_a=150$ active users. Performance baselines in comparison are conventional EM-AMP \cite{EM-AMP1,AMP3}, SOMP+least square \cite{baseline2} and AoA codebook-based \cite{baseline1}.}\label{sim:Performance_N_o}
\end{figure}
\subsubsection{Performance versus Active Ports Number $N_o$}
Fig.~\ref{sim:Performance_N_o} compares algorithm performance and complexity for varying \(N_o\), with 150 active users and \(5\) dB SNR. The EM-AMP framework excels in low complexity, reducing computational overhead by 53.46\% (proposed) to 85.03\% (conventional) compared to least squares at $N_o=62$. After \(N_o=6\), the proposed scheme costs much less complexity but the performance loss is marginal. Moreover, angular information is critical for FAS CSI acquisition. AoA codebook-based methods lack precision with few active ports due to low resolution but achieve the lowest NMSE with sufficient ports due to adequate angle domain resolution.
\section{Conclusion}\label{sec.5}
We introduce an EM-AMP framework for CSI acquisition in FAS, with update rules using geographical signal priors, enhancing estimation precision and convergence. The EM-AMP framework cut complexity by 50\%--85\% in large-scale deployments with minimal performance loss versus existing methods. Overall, AMP-based solutions for FAS provide low-complexity, feasible, and flexible CSI acquisition.
\section*{Acknowledgment}

The work of Z. Zhang, J. Dang and Z. Zhang is supported in part by NSFC (61971136, 61960206005), the Fundamental Research Funds for the Central Universities (2242022k60001, 2242021R41149, 2242023K5003). The work of D. Morales-Jimenez is supported in part by the State Research Agency (AEI) of Spain.

\end{document}